# Torque on slowly moving electric or magnetic dipoles in vacuo

G. Asti and R. Coïsson

Department of Mathematics, Physics and Informatics
University of Parma, 43100 Parma, Italy

e-mails: giovanni.asti@fis.unipr.it, roberto.coisson@fis.unipr.it

**Abstract**
The torque on a moving electric or magnetic dipole in slow motion is deduced using the Lorentz transformation of the fields to first order in v/c. It is shown that the obtained equations are independent of the model adopted for the dipole, whether it is of Amperian or Gilbertian type, thus showing the complete validity of the Ampère equivalence principle even in dynamical conditions. The torque is made of three terms: beside that due to the direct torque on the dipole there are two more terms: one due to the torque on the associated perpendicular dual-dipole caused by motion, while the other is the inertial torque due to the displacement of the dipole which carries with it the field linear momentum, or the hidden momentum.

The calculation of the torque on a moving electric (ED) or magnetic dipole (MD) to first order in v/c has been the subject of several papers [1-4], in particular in connection with electromagnetic momentum and "hidden momentum". Here we want to show that a straightforward application of relativistic transformation of fields, as in our previous paper where forces on moving dipoles were calculated [5], allows a clear physical interpretation without specific hypotheses or models, and naturally makes evident the Ampère equivalence principle.
The aim of the calculation is to obtain the torque in the approximation of first order in v/c.

Consider first the moving MD **m'** in its proper frame S'; then the torque is

$$\mathbf{N}_{md} = \mathbf{m}' \times \mathbf{B}' \tag{1}$$

The relativistic transformation of fields to the laboratory frame S are :

$$\mathbf{B}' = \mathbf{B} - \frac{1}{c^2}\mathbf{v} \times \mathbf{E} \qquad \mathbf{E}' = \mathbf{E} + \mathbf{v} \times \mathbf{B} \tag{2}$$

The substitution of the field gives the torque equation in frame S:

$$\mathbf{N}_{md} = \mathbf{m} \times (\mathbf{B} - \frac{1}{c^2}\mathbf{v} \times \mathbf{E}) = \mathbf{m} \times \mathbf{B} - \frac{1}{c^2}\mathbf{m} \times (\mathbf{v} \times \mathbf{E}) \tag{3}$$

But $\quad \mathbf{m} \times (\mathbf{v} \times \mathbf{E}) = -\mathbf{v} \times (\mathbf{E} \times \mathbf{m}) - \mathbf{E} \times (\mathbf{m} \times \mathbf{v})$. (4)

Then

$$\mathbf{N}_{md} = \mathbf{m} \times (\mathbf{B} - \frac{1}{c^2}\mathbf{v} \times \mathbf{E}) = \mathbf{m} \times \mathbf{B} + \frac{1}{c^2}[\mathbf{v} \times (\mathbf{E} \times \mathbf{m}) + \mathbf{E} \times (\mathbf{m} \times \mathbf{v})] \tag{5}$$

Finally, if we define $\mathbf{p}_v = \frac{1}{c^2} \mathbf{v} \times \mathbf{m}$ the apparent (or joint) electric dipole due to motion, we have

$$\mathbf{N}_{md} = \mathbf{m} \times \mathbf{B} + \mathbf{p}_v \times \mathbf{E} + \frac{1}{c^2} \mathbf{v} \times (\mathbf{E} \times \mathbf{m}) \quad . \tag{6}$$

Consider now the ED **p.** The torque in the proper frame is

$$\mathbf{N}_{ed} = \mathbf{p}' \times \mathbf{E}' = \mathbf{p} \times \mathbf{E} + \mathbf{p} \times (\mathbf{v} \times \mathbf{B}) \quad , \tag{7}$$

but $\quad \mathbf{p} \times (\mathbf{v} \times \mathbf{B}) = -\mathbf{v} \times (\mathbf{B} \times \mathbf{p}) - \mathbf{B} \times (\mathbf{p} \times \mathbf{v}) \quad ,$ (8)

and $\mathbf{m}_v = \mathbf{p} \times \mathbf{v}$ is the joint magnetic dipole due to motion,
so the transformed torque results

$$\mathbf{N}_{ed} = \mathbf{p}' \times \mathbf{E}' = \mathbf{p} \times \mathbf{E} + \mathbf{m}_v \times \mathbf{B} + \mathbf{v} \times (\mathbf{p} \times \mathbf{B}) , \tag{9}$$

where the first term is the torque on the ED due to E, the second one the torque on the joint MD due to **B**, and the third one is -d/dt **r** x (**B** x **p**) that is the inertial torque due to time variation of the angular momentum, which is caused by the translation of the field linear momentum, **B** x **p**.

This implies that **B** x **p** is a momentum (as **E**x**m** in the case of MD). What kind of momentum depends on the model used: electromagnetic momentum for dipoles modeled with charges (electric or magnetic), otherwise as hidden momentum.
We remark that there is no need to especially introduce the apparent (or "joint") dipole due to motion: it comes out automatically [6].

The above calculation shows once more the perfect symmetry between magnetic and electric dipoles, independently on the adopted models, confirming the extended validity of the Ampère equivalence principle to moving dipoles *in vacuo*.

In the case of time-varying dipole moments, the same approach would give the same results.

The strange story of the "missing torque" is probably due to a forgotten explanation based of the inertial torque due to time variation of the angular momentum caused by the presence of hidden momentum [7] (as mentioned above for the interpretation of the 3rd term). We say here *"forgotten explanation"* because a similar argument, based on the inertial torque, was used for the interpretation of the Trouton-Noble historical experiment. [8]

*A particular example*
This example makes evident the importance of the 3rd term.
An ED lying in the x'y' plane is oriented at an angle θ with respect to x-axis and is only under the action of a uniform magnetic field **B'** parallel to z'-axis. The torque on the stationary ED is

$$\mathbf{N}'_{ed} = \mathbf{p}' \times \mathbf{E}' = 0 \tag{10}$$

being **E'**=0.

In the laboratory frame the dipole travels along the positive x-axis at velocity v. So

$$\mathbf{E} = \mathbf{E'} - \mathbf{v} \times \mathbf{B'} = -\mathbf{v} \times \mathbf{B'} \quad \text{parallel to y} \; ; \quad \mathbf{B} = \mathbf{B'} + \frac{1}{c^2}\mathbf{v} \times \mathbf{E'} = \mathbf{B'} \quad \text{parallel to z} \tag{11}$$

$$\mathbf{m}_v = \mathbf{p} \times \mathbf{v} \quad \text{parallel to z.}$$

From eq. 9 we have

$$\mathbf{N}_{ed} = \mathbf{p'} \times \mathbf{E'} = \mathbf{p} \times \mathbf{E} + \mathbf{m}_v \times \mathbf{B} + \mathbf{v} \times (\mathbf{p} \times \mathbf{B}) = -\mathbf{p} \times (\mathbf{v} \times \mathbf{B'}) + (\mathbf{p} \times \mathbf{v}) \times \mathbf{B'} + \mathbf{v} \times (\mathbf{p} \times \mathbf{B'}) \tag{12}$$

obviously equal to zero, as it should be. But $\mathbf{m}_v$ is parallel to **B.** Hence the torque due to **E** is entirely balanced by the inertial term due to to the field momentum **p** x **B**.